\documentclass[journal=jacsat,manuscript=article,layout=twocolumn]{achemso}
\setkeys{acs}{articletitle = true}

\usepackage[version=3]{mhchem} % Formula subscripts using \ce{}
\usepackage{color}
\usepackage{comment}
\usepackage[utf8]{inputenc} 
\usepackage{mathtools}
\usepackage{comment}
\usepackage{afterpage}
\definecolor{kagome}{RGB}{204,0,0}
\definecolor{dense}{RGB}{43,178,76}
\definecolor{fluid}{RGB}{44,138,236}

\author{Stewart A. Mallory}
\email{smallory@caltech.edu}
\affiliation[Caltech]
{Division of Chemistry and Chemical Engineering, California Institute of Technology, Pasadena, California 91125, United States}

\author{Angelo Cacciuto}
\email{ac2822@columbia.edu}
\affiliation[Columbia University]
{Department of Chemistry, Columbia University, New York, New York 10027, United States}

\title[Activity-Enhanced Self-Assembly of a Colloidal Kagome Lattice]{Activity-Enhanced Self-Assembly of a Colloidal Kagome Lattice}

\begin{document}

\begin{tocentry}

\centering 
\includegraphics[]{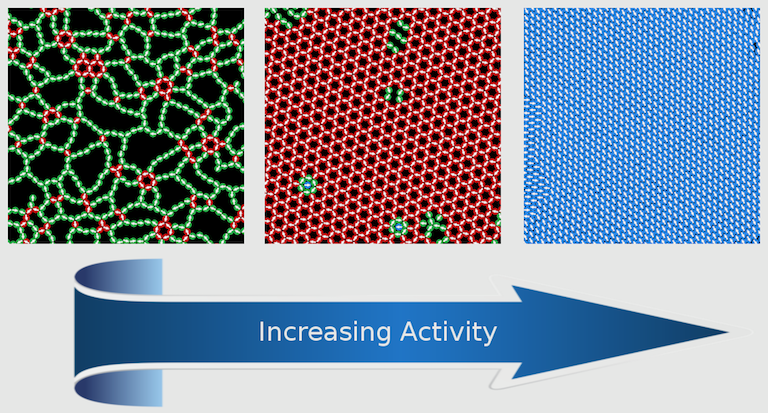}

\end{tocentry}

\begin{abstract}
Here, we describe a method for the enhanced self-assembly of triblock Janus colloids targeted to form a kagome lattice. Using computer simulations, we demonstrate that the formation of this elusive structure can be significantly improved by self-propelling or \textit{activating} the colloids along the axis connecting their hydrophobic hemispheres. The {process} by which metastable aggregates are destabilized and transformed into the favored kagome lattice is quite general, { and we argue this \textit{active} approach provides a systematic pathway to improving the self-assembly of a large number of colloidal structures.}
\end{abstract}

\section{Introduction}

The current rise in popularity of colloidal active matter is due in large part to the recent advances in colloidal synthesis. Through the pioneering work of synthetic chemists and material scientists, there now exists a large number of experimental realizations of synthetic microswimmers and active colloids. \cite{Wang2015,Zhang2017,ebbens_active_2016,Patteson2016,Gibbs2011,dey_chemically_2017,Bechinger2016,romanczuk_active_2012} These \textit{active particles} can be thought of as the synthetic analogues of swimming bacteria. However, a major benefit of these synthetic variants over their biological counterpart is the ability to systematically tailor interparticle interactions and dynamically modulate swim speed. \cite{gao_dynamic_2017,Singh2017,Lin2017,Fournier-Bidoz2005,Paxton2005,Palacci2014,kaiser_active_2015,Theurkauff2012,pohl_dynamic_2014,Jiang2010,Wang2013,Ahmed2014,Gao2013} The functionality of synthetic active particles potentially makes them the ideal tool to manipulate and self-assemble matter at the microscale.\cite{solovev_self-propelled_2012,Mallory2018,Mallory2017,Shklarsh2012,Gao2013a,Palacci2013,DiLeonardo2016,Wensink2014,Spellings2015,hasnain_crystallization_2017,wykes_dynamic_2016,Zhang2016,Gao2014} 

In a recent {theoretical study}, \cite{Mallory2016} it was {suggested} that active forces can dramatically improve the self-assembly of colloidal building blocks designed to form capsid-like structures. The key design principle for these small compact structures, {which has been experimentally demonstrated in systems of gliding microfilaments, \cite{Hess2006,Hess2005}} was to have the self-propelling forces of the colloids sum close to zero within the target structure. This generates a unique environment where the target structure is further stabilized by active forces, and any competing metastable structures are quickly destroyed. This approach is successful in leveraging activity to improve the self-assembly of small compact colloidal structures; however, the problem becomes less obvious when considering the self-assembly of macroscopically large crystals or periodic structures. In this work, we explore to what extent activity can be used as an extra handle to promote the self-assembly of colloidal building blocks designed to form stable periodic structures and under what conditions the complex self-assembly landscape can be simplified and biased toward a particular target structure. 

\begin{figure}[t!]
 \centering
    \includegraphics[width=0.45\textwidth]{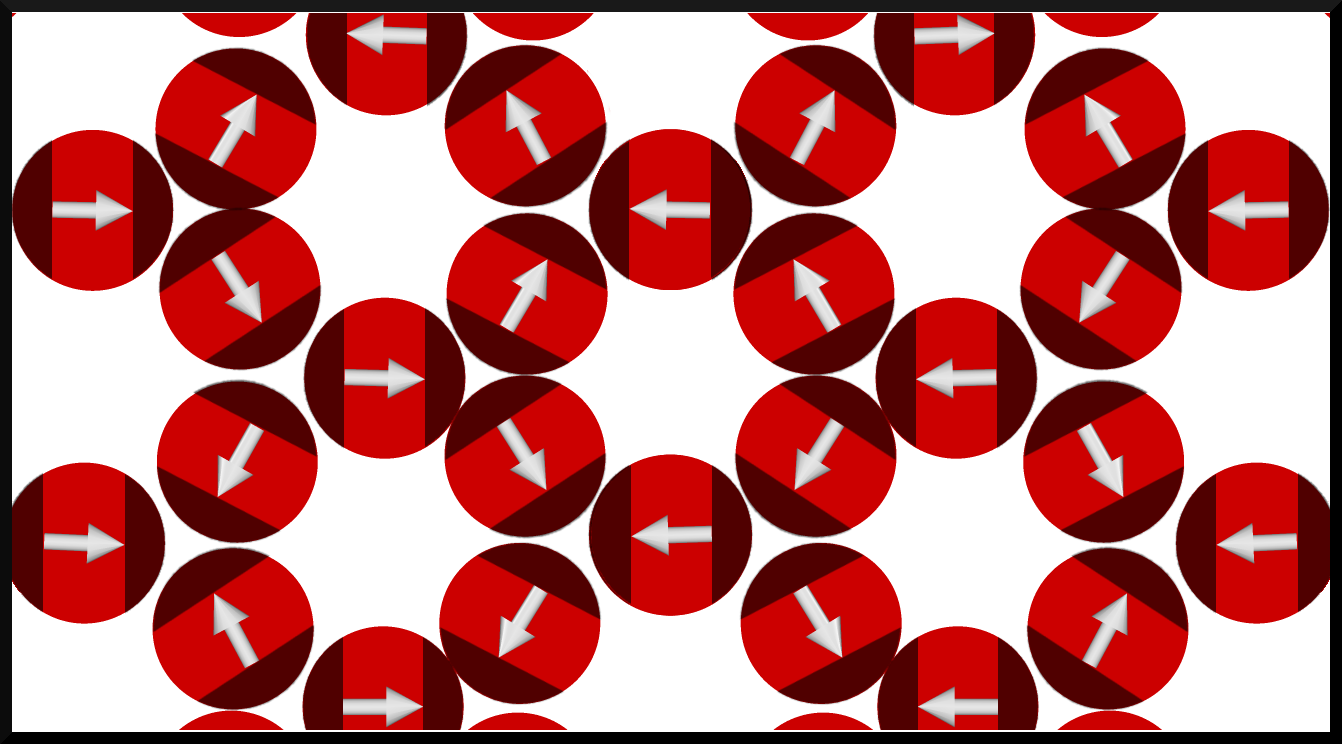}
  \caption{Triblock Janus colloids tailored to self-assembly into a kagome lattice. There exist a short-ranged attraction between hydrophobic patches (black), while the equatorial regions (red) contribute only volume exclusion interactions. The white arrows indicate the direction of self-propulsion, which is fixed and colinear with the line connecting the center of the hydrophobic patches.}
  \label{fig:schematic}
\end{figure}

As our model system, we consider the heavily studied experimental system of triblock Janus colloids. \cite{Mao2013,Chen2011DirectedLattice,Chen2011,Zhang2015,Reinhart2018,Reinhart2016EquilibriumColloids} As shown in Figure \ref{fig:schematic}, the triblock system is designed to self-assemble into a kagome lattice. We have selected this particular colloidal system for two important reasons. The first being that the kagome lattice, and more generally open-cell colloidal crystals, are of particular interest because of their novel optical and mechanical properties. \cite{Galisteo-Lopez2011,Chapela2014a,Desiraju2007,Souslov2009a,su2012hierarchically}

The second is that open-cell lattices are notoriously difficult to self-assemble and present a unique challenge within the materials community. \cite{shin_theory_2014,Sacanna2011,chen_janus_2012,Morphew2018,bianchi_phase_2006,Yi2013,Rocklin2014} For most {patchy} colloidal systems, including the triblock system, there exist an ensemble of extended structures that directly compete with the formation of the target structure. And if the self-assembly pathway is not fully understood and system conditions are not highly optimized, the time required to form the target structure can become prohibitively long. \cite{whitelam_statistical_2015,whitelam_role_2009,Wilber2007,Miller2009,Jankowski2012}For the triblock system, the formation of the kagome lattice is observed to be a two-step process. A fast first step that involves the formation of a metastable web-like network or gel followed by a second significantly slower step.  The slow second step involves local structural rearrangements that morph the kinetically favored gel state into the thermodynamically favored kagome lattice. This multistep mechanism is common in complex self-assembly processes that rely on {patchy} particles interacting through short-range interactions as their colloidal building blocks. Thus, the viability of a large class of self-assembled materials hinges on the kinetics of the second step, which in turn is determined by the robustness of thermal fluctuations and their ability to rearrange colloids from the disordered metastable state to the thermodynamically preferred target structure.   

The premise of our approach is built around finding alternative and more efficient self-assembly pathways by leveraging the unique driven nature of active colloids. By introducing activity as a new dimension in the self-assembly parameter space, we demonstrate how this new control variable makes it possible to bias the selection of a particular crystal structure against the formation of unwanted metastable states and greatly improves the overall yield and rate of self-assembly in the triblock system. 

\section{Model and Analysis}

Each triblock Janus colloid is modeled as a sphere of diameter $\sigma$ and undergoes Brownian dynamics at a constant temperature $T$ according to the coupled overdamped equations: 

\begin{equation}
\dot{\pmb{r}}(t) = U \, \pmb{n}(t) + \sqrt{2D}\,\pmb{\xi}(t) + \gamma^{-1} \pmb{F}\{r_{ij}\} 
\end{equation}

\begin{equation}
\dot{\pmb{{n}}}(t)=  \sqrt{2D_R}\, \pmb{\xi}_R(t) \times \pmb{n}(t) + \gamma_R^{-1}\pmb{T}\{r_{ij}\}
\end{equation}

\noindent  where the translational diffusion of the colloid is given by the Stokes--Einstein relation $D=k_{\rm B}T\gamma^{-1}$, with $\gamma$ being the translational friction due to the fluid. For spherical colloids in a low Reynolds number environment, it is reasonable to assume rotational diffusion is thermal in origin and given by $D_R=k_{\rm B}T\gamma_R^{-1}=3D\sigma^{-2}$, where the rotational friction is given by $\gamma_R$. In this study, we prefer not to specify a particular propulsion mechanism and instead generally assume that each triblock Janus colloid swims at a fixed speed $U$ along a predefined orientation unit vector $\pmb{n}(t)$. {In this work, we also neglect all hydrodynamic interactions between colloids. Given the relatively small swim speeds involved in this study, it is reasonable to assume that long-range hydrodynamics interactions will be relatively weak and will only quantitatively change the self-assembly behavior of the triblock Janus system.}

The two hydrophobic patches that give the colloids their triblock nature are located at opposite poles of the colloid, with each hydrophobic patch accounting for $\approx 28\%$ of the particle surface area. The polar equatorial region accounts for the remaining $44\%$ of the surface area. A key requisite of this approach is that the swimming axis of the colloid is aligned with the center-line of the two hydrophobic hemispheres as shown in Figure \ref{fig:schematic}. By choosing a propelling axis for the colloids that is colinear with the center-line of the two hydrophobic hemispheres, we minimize internal stresses and shearing forces within the kagome structure. The random Brownian forces and torques acting on each colloid are modeled as Gaussian white-noise and are characterized by $\langle \pmb{\xi}(t)\rangle = 0$ and $\langle \xi_\alpha(t) \xi_\beta(t^\prime)\rangle = \delta_{\alpha \beta}\delta(t-t^\prime)$ and $\langle \pmb{\xi}_R(t)\rangle = 0$ and $\langle \xi_{R \alpha}(t) \xi_{R \beta}(t^\prime)\rangle = \delta_{\alpha \beta}\delta(t-t^\prime)$, respectively. The interparticle forces and torques that arise from the chemical patterning of the colloid's surface are given by $\pmb{F}\{r_{ij}\}$ and  $\pmb{T}\{r_{ij}\}$, respectively, with $\alpha,\beta \in \{x,y\}$. The interparticle pair potential for the triblock Janus colloids was implemented such that the phenomenology agrees with the experimental work of \citeauthor{Chen2011DirectedLattice} \cite{Chen2011DirectedLattice} and the extensive numerical simulations of Janus colloids in the literature. \cite{Hong2008ClustersSpheres,Liang2006ClustersSpheres,Miller2009,Mallory2017} {Additionally, our model exhibits qualitatively similar phase behavior to previous thermodynamic studies on equilibrium triblock Janus colloids. \cite{Reinhart2016EquilibriumColloids,Reinhart2018,Romano2011,Rocklin2014}} A {more complete} discussion of this {patchy} potential and its functional form can be found in the Supporting Information. Below we list the key features of the anisotropic interaction potential, which drives the formation of the kagome lattice:

\begin{itemize}
    \item Experimentally, self-assembly is induced by the addition of salt, which screens long-ranged electrostatics and allows hydrophobic attractions to dominate. In simulations, this is manifested by a short-range attraction between the hydrophobic patches, which we often refer to as a ``bond", and is characterized by a binding energy $\varepsilon$, which is cutoff at a separation distance of $1.5 \sigma$.
    \item Due to the electrostatic screening, we assume that the equatorial polar regions can effectively be described by a short-ranged volume excluding potential, which is cutoff at $2^{1/6}\sigma$.
    \item The angular potential between the colloids is rather flat for a range of angular displacements.
     The only nonthermal torques are those induced at the boundary between the polar and hydrophobic region of the colloids. This is consistent with experimental observations that show a large degree of rotational freedom when two particles are in contact with each other with their hydrophobic patches facing each other.
\end{itemize}

In this work, all simulations contain $N=10,000$ triblock Janus colloid, which are strictly confined to self-propel in two dimensions. Both the motion and the orientation vectors of the colloids are confined to evolve in the xy-plane. This mimics the reference experimental system where self-assembly is templated to occur on a surface after sedimentation. The length and energy scales were chosen to be $\sigma$ and $k_{\rm B}T$, respectively, while $\tau=\sigma^2D^{-1}$ is the unit of time. All simulations were run for $10^4\tau$ with a time step of $\Delta t=10^{-5}\tau$. {All results are reported in reduced Lennard-Jones units. As a useful reference, an active colloid in our simulations with swim speed $U=1$ in reduced Lennard-Jones units corresponds to a swim speed of $\approx 1 \  \mu$m/sec for an active colloid with a diameter of $\approx 1 \  \mu$m.}

Three different volume fractions were considered: $\phi=0.2$, $\phi=0.4$, and $\phi=0.6$. For each volume fraction, a range of temperatures $T$ and self-propelling speeds $U$ were considered such that this study spans the parameter space where formation of the kagome lattice is possible. Both the size of the hydrophobic patches and the magnitude of the binding energy ($\varepsilon= 15 k_{\rm B}T$) were tuned to maximize the self-assembly of the kagome lattice in the equilibrium system $U=0$ at a temperature of $T=1$ and volume fraction $\phi=0.6$. All simulations were carried out using a modified version of the HOOMD-Blue simulation package, \cite{Glaser2015,Anderson2008} and a majority of the analysis was done within the OVITO visualization and analysis package. \cite{Stukowski2010}

Using a shape-matching approach described by \citeauthor{Keys2011}, \cite{Keys2011} the success or {yield} of self-assembly is quantified in terms of local structural features. In general, a colloid is identified as part of a given structural phase by analyzing the nearest neighbors within a cutoff of $1.25 \sigma$ and the relative orientation of those neighbors around the particle. Our approach is equivalent to the method employed by \citeauthor{Chen2011DirectedLattice} \cite{Chen2011DirectedLattice} to identify kagome order in their experimental work. The percent yield of a phase is simply defined as the percentage of colloids of a given structural type, and we delineate between three structural phases: kagome, hexagonal, and a disordered fluid phase.  A more thorough discussion on phase identification for patchy colloidal systems is provided in the Supporting Information.

\section{Results and Discussion}

\begin{figure}[t!]
 \centering
    \includegraphics[width=0.45\textwidth]{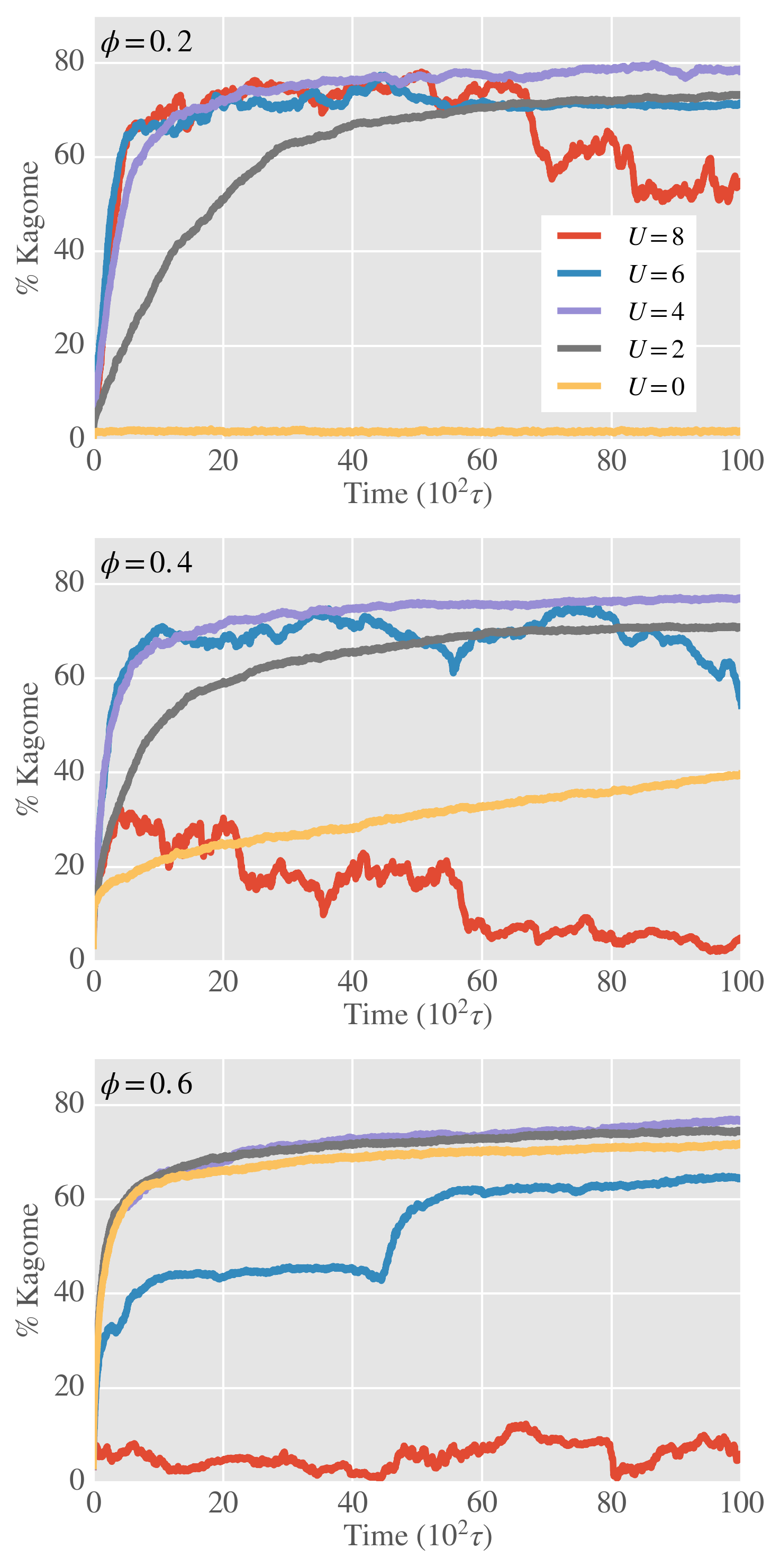}
  \caption{Percent yield of the kagome lattice as a function of time for three different volume fractions $\phi=0.2$, $\phi=0.4$, and $\phi=0.6$. The system temperature was fixed at $T=1$ for all volume fractions and propelling speeds reported in the figure.}
  \label{fig:time_trace}
\end{figure}

Figure \ref{fig:time_trace} illustrates how introducing activity as a control variable can improve the self-assembly of patchy colloidal systems. Here, we present the percent yield of the kagome lattice as a function of time for the three different volume fractions considered in this study. In Figure \ref{fig:time_trace}, the temperature is fixed at $T=1$ while a range of self-propelling speeds were explored. Both the active ($U>0$) and equilibrium ($U=0$) triblock systems exhibit the characteristic two-step self-assembly growth dynamics. There is an initial burst of kagome order followed by a regime of slow but constant growth. At equilibrium, the overall yield is heavily influenced by the bulk volume fraction. For dilute systems (e.g., $\phi=0.2$), the initial jump in kagome order is small $(<10\%)$ and the rate of self-assembly is nearly zero {(Movie S1)}. The slow rate of self-assembly is due predominately to the dilute nature of the system. For the triblock system at equilibrium, random thermal fluctuations are solely responsible for breaking apart malformed aggregates and regulating diffusion across the system. So as the system becomes increasing dilute, it inherently takes more time for a colloid to detach, reorient, and diffuse to a region where it can be properly integrated into the kagome lattice.

\begin{figure}[t!]
 \centering
    \includegraphics[width=0.45\textwidth]{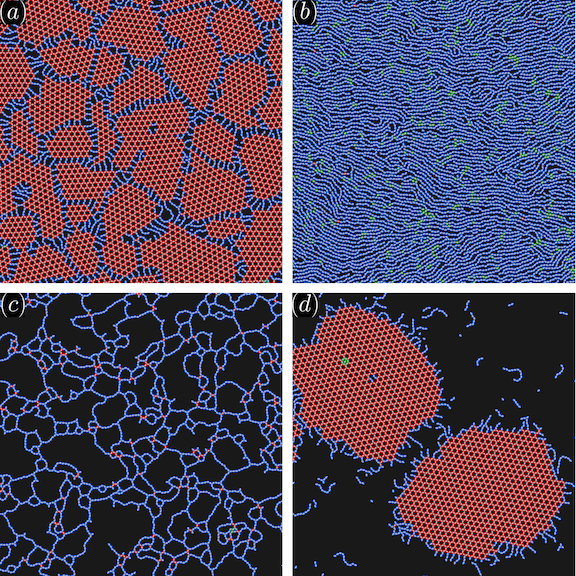}
 \caption{Representative snapshots of the triblock Janus system for different state points $(\phi,U,T)$. (a) The best performing equilibrium system at state point $(\phi=0.6,U=0,T=1.2)$. (b) Disordered fluid phase at high temperature ($T=1.2$) for equilibrium system at volume fraction $\phi=0.6$. (c) Representative snapshot of metastable web-like network at state point $(\phi=0.2,U=0,T=0.2)$. (d) The best performing active system at state point $(\phi=0.2,U=4,T=1.2)$.
 }
  \label{fig:equil_snapshots}
\end{figure}

At the opposite end of the spectrum, if the bulk volume fraction is too high $(\phi\gg0.6)$, the self-assembly dynamics becomes sluggish, and the system is prone to become trapped in a metastable disordered state (data not shown). There exists a window in volume fraction where self-assembly of the kagome lattice is favored, and we empirically find that $\phi=0.6$ is an optimal volume fraction to maximize the rate of self-assembly for the  equilibrium system. At this particular volume fraction, the colloids almost immediately find the proper coordination and orientation to form the kagome lattice. Self-assembly at this volume fraction also leads to highly polycrystalline structures, as formation of the kagome lattice occurs nearly simultaneously throughout the sample (see Figure \ref{fig:equil_snapshots}a). The simulations at volume fraction $\phi=0.6$ predominately serve as a reference and provide insight into the effects of activity for an optimized equilibrium self-assembly process. 

Interestingly, the percent yield of the kagome lattice increases significantly once a small amount of activity is introduced. The self-propulsion of the triblock colloids generates a unique nonequilibrium environment that constantly allows for rearrangements of any mechanically unstable structure and guides the formation of the thermodynamically and mechanically stable kagome lattice {(see Movie S2)}. For the two smaller volume fractions $\phi=0.2$ and $\phi=0.4$, the rate of self-assembly is dramatically faster than the equilibrium system. As expected, the improvement in self-assembly is less dramatic for the highly optimized equilibrium system at $\phi=0.6$. It is also important to note from Figure \ref{fig:time_trace} that the percent yield has a nonmonotonic dependence on the self-propelling speed. This suggests that a balance must be struck between active and adhesive forces to maximize the yield, or more generally that activity cannot be implemented blindly, but rather has to be done in a manner that is commensurate with the target structure and the other system parameters (density, temperature, binding energy, etc.). 

\begin{figure}[t!]
 \centering
    \includegraphics[width=0.45\textwidth]{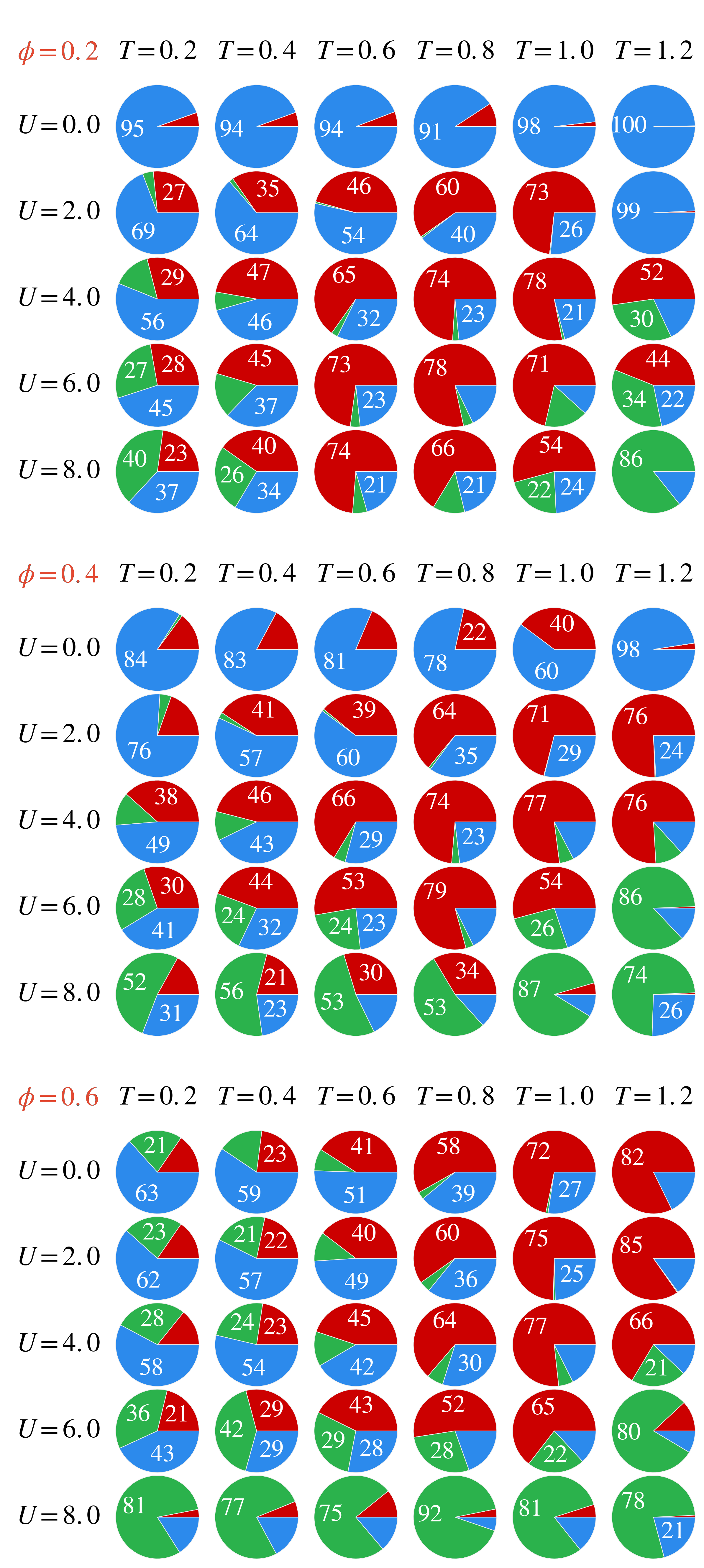}
  \caption{Percentage of kagome {(red)}, hexagonal {{dense} (green)}, or disorder phase {{fluid} (blue)} for the three volume fractions (from top to bottom) $\phi=0.2$, $\phi=0.4$, and $\phi=0.6$. The numerical value of the corresponding yield is indicated at the center of each slice (yields below $20\%$ are unlabeled).}
  \label{fig:yield}
\end{figure}

Figure \ref{fig:yield} presents the phase compositions as a function of temperature $T$ and self-propelling speed $U$ after allowing the system to evolve for $10^4\tau$. For volume fractions $\phi=0.2$ and $\phi=0.4$ the equilibrium system does not exhibit a significant degree of crystalline order for any of the temperatures considered in this study. The one exception is for the intermediate volume fraction $\phi=0.4$, where there is a single temperature $T=1.0$ where the formation of the kagome lattice is appreciable $(40\%)$, albeit the rate of self-assembly is rather slow (see Figure \ref{fig:time_trace}). For $\phi=0.6$, the equilibrium system exhibits a significant degree of kagome order for $T=1.0$ ($>70\%$ yield) and $T=1.2$ ($>80\%$ yield). A snapshot of the final configuration of the best performing equilibrium system $T=1.2$ is given in Figure \ref{fig:equil_snapshots}a. For $\phi=0.2$, the equilibrium triblock system assembles into a porous percolating networks (gel) at low temperatures (see Figure \ref{fig:equil_snapshots}(c)). For $\phi=0.6$ at $T\leq 0.8$, the system also forms a gel network; however, due to the increase in volume fraction, there are a few small regions having hexagonal and kagome compatible crystalline order.  The branching factor of these networks is inversely proportional to temperature. As temperature is increased, we observe the formation of linear chains with minimal branching and predominately a single bond per patch. For volume fraction $\phi=0.2$ at $T=1.2$ the system forms a disordered dynamic fluid of polydisperse flexible rods. At larger volume fractions $\phi=0.4$ and $\phi=0.6$ at $T=1.2$ these flexible rods stack and wind around one another (see Figure \ref{fig:equil_snapshots}b) forming patterns similar to fingerprint textures observed in liquid crystals. \cite{Gennes1993} 

Once activity is introduced in the system it is always possible to find a range of temperatures and activities for which a high yield of kagome assembly can be obtained. Remarkably, this is true across all three volume fractions considered in this study. A snapshot of the final configuration of the active system at state point $(\phi=0.4,U=4,T=1.2)$ is given in Figure \ref{fig:equil_snapshots}d. The physical mechanism that leads to this level of control over local structure can be attributed to two unique features of active systems. The first is giant long-lived density fluctuations. These fluctuations are closely related to the heavily studied phenomenon of motility induced phase separation and are generally believed to be a generic feature of active systems that are driven out of equilibrium by a persistent local energy input that breaks detailed balance.  

For the best performing region of the active triblock system, we observe the kagome lattice form out of local high density fluctuations that develop from frequent collisions among the self-propelled linear precursors. These large fluctuations in density are especially important early on in the self-assembly process. For many self-assembly processes, the growth of a particular target structure requires the formation of a critical nucleus of a characteristic size. By controlling the level of activity in the system, it is possible to modulate the magnitude and lifetime of density fluctuations. The fact that activity alone can generate giant fluctuations in density has important implications for colloidal self-assembly and is a key ingredient in developing our approach to {active} self-assembly. In addition, self-propulsion will in general improve the kinetics of self-assembly, as the dynamics becomes inherently faster.  The rate at which an unbounded colloid can explore the system is now set by the swim speed $U$, and is no longer solely regulated by the temperature $T$. The enhanced diffusivity scales as $D\sim U^2/D_r$ in the dilute limit. \cite{TenHagen2009}

The second contribution specifically targets those colloids bound in the percolating network, it is more important later on in the self-assembly process, and is central to healing defects and grain boundaries.  A unique aspect of patchy colloids including the triblock system is that they typically do not form plastic crystals. The rotation of each colloid is severely hindered when bonded to its neighbors. When the system is active, the colloids in the crystal are orientationally locked and exert a force $F \sim \gamma^{-1} U$ along the direction of their propelling axis.  Any extended structure developing in the system that competes with the formation of the kagome lattice, such as the metastable gel state pictured in Figure \ref{fig:equil_snapshots}c, become mechanically unstable upon the introduction of active forces. This structural imbalance gives rises to internal stresses and shearing forces which drive the compaction of the low density network into the more stable kagome lattice. This compaction and reorganization process will continue until all colloids have found a compatible orientation within the kagome lattice. However, if the self-propelling speed becomes too large this effect can contribute to the partial collapse of the kagome lattice into one with hexagonal symmetry. 

Interestingly, we observe extended regions of coexistence between kagome and hexagonal crystals with the latter crystal structure becoming more stable at large activities. {This is consistent with the notion that an increase in activity corresponds to an increase in swim pressure, \cite{Mallory2014,Takatori2014} which for sufficiently large activities drives the collapse of the kagome lattice. However, it should be appreciated that the nature of pressure in active systems is much richer than its equilibrium counterpart. A recently published study provides a more rigorous framework for contextualizing how pressure can be manipulated in active system to drive colloidal self-assembly. \cite{Omar2019} Lastly, it should be noted that in the equilibrium system the hexagonal phase is expected {only for very low temperatures} where attractive interactions dominate over thermal forces and at larger volume fractions where hexagonal order is entropically favored. \cite{Mao2013}} Thus, the formation and coexistence of a hexagonal phase under these conditions is an inherent non-equilibrium property of the active triblock system.

\begin{figure}[t!]
 \centering
    \includegraphics[width=0.45\textwidth]{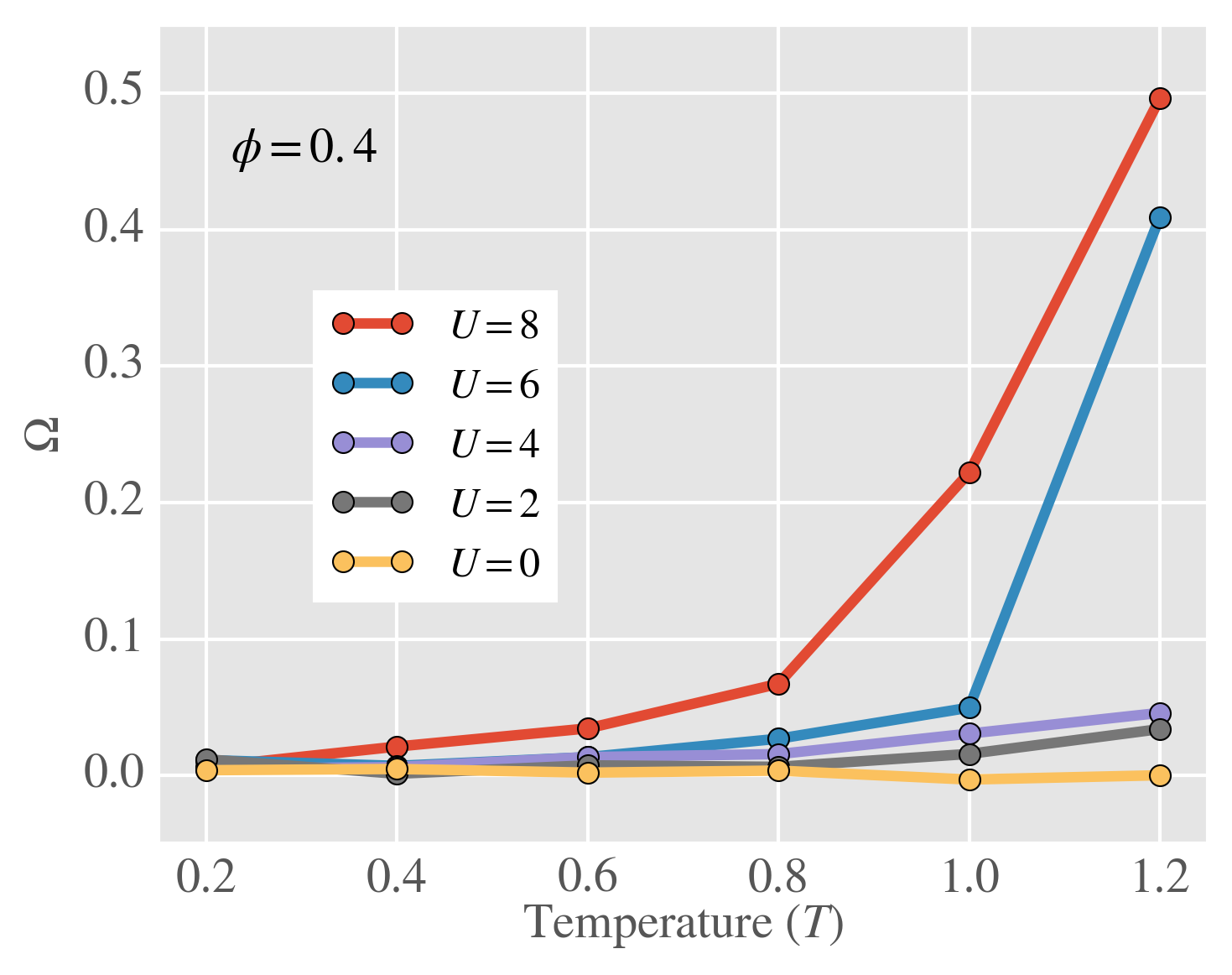}
  \caption{Global polar order parameter $\Omega$ as a function of temperature $T$ for various propelling speeds $U$ at volume fraction $\phi=0.4$. Active systems at high temperatures exhibit a significant degree of orientational alignment, while at low temperature colloids are predominately randomly oriented.
  }
  \label{fig:align}
\end{figure}

\begin{figure}[t!]
 \centering
    \includegraphics[width=0.45\textwidth]{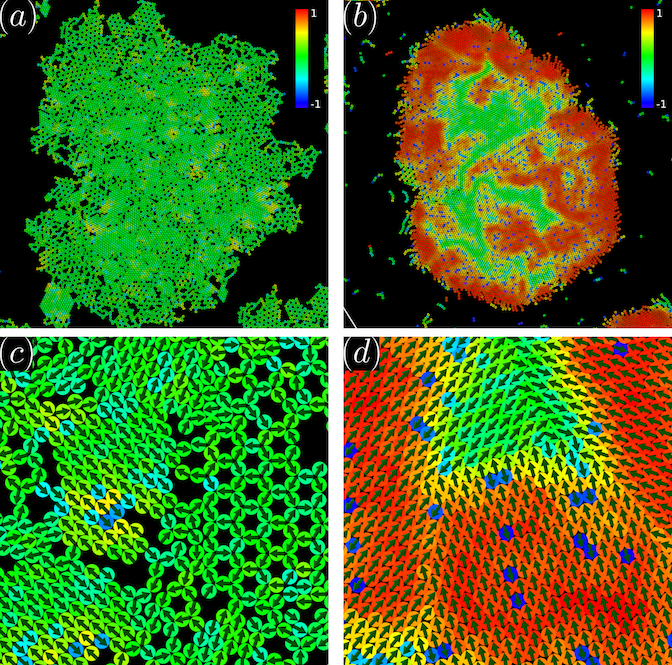}
  \caption{Snapshot of final configuration for active self-assembly at low (a) and high temperature (b). The self-propelling speed is fixed at $U=8$ and $\phi=0.4$. The coloring of each colloid is given by the local alignment $\omega_i$ as defined in Eq. 3. (a) At low temperature ($T=0.2$), we observe coexistence between kagome and hexagonal structures. Within both crystal structures, the colloids exhibit low to no alignment. (b) At high temperatures ($T=1.2$), colloids within the center of the cluster exhibit very little polar alignment (green), while colloids on the perimeter exhibit a high degree of alignment $(red)$. There is a small fraction of colloids within these order domains that are antialigned with their neighbors (blue). Panels (c) and (d) present magnified views of (a) and (b), respectively.}
  \label{fig:align_snapshot}
\end{figure}

{Unexpectedly}, there are two different regimes where hexagonal order naturally develops in the active system, and although the corresponding structures both exhibit translational hexagonal symmetry, they differ significantly in the orientation of the triblock colloids within the lattice. {The two different flavors of hexagonal order in the active system are shown in Movies S3 and S4.} We should stress that the formation of a stable hexagonal lattice in both the active and equilibrium system requires a global realignment of the  orientations of the colloids within the lattice. As shown in Figure \ref{fig:align_snapshot}c and Figure \ref{fig:align_snapshot}d, colloids within the hexagonal lattice are oriented such that their orientation axes are parallel to their neighbors. This configuration maximizes the interaction between hydrophobic patches at large densities and is different than that associated  with the kagome lattice (see Figure \ref{fig:schematic}) where the overlap between hydrophobic patches is maximized when the angle between orientation vectors is $60^o$. This alignment of the orientation vectors gives rise to a high level of nematic order within the hexagonal lattice. If we now take into account the direction of the propelling forces of the triblock colloids (i.e. polar order) we find a substantial difference in the orientational distribution for self-assembled structures at low and high temperature. For this purpose, we introduce a global polar order parameter $\Omega$ defined as 

\begin{equation}
\Omega = \frac{1}{N} \sum_{i=1}^N \omega_i =\frac{1}{NM_i} \sum_{i=1}^N\sum_{j=1}^{M_i} \pmb{n}_i\cdot \pmb{n}_{j}
\end{equation}

\noindent where $\pmb{n}_i$ is the orientation of the propelling vector of the $ith$ colloid and $M_i$ is its corresponding number of neighbors within a cutoff distance $r_c=3\sigma$. Figure \ref{fig:align} shows $\Omega$ as a function of $T$ for different values of $U$ measured from the final configurations at volume fraction $\phi=0.4$. If each colloid has its axis of propulsion oriented in the same direction as its neighbors, then $\Omega=1$ and if the colloidal propelling axes are randomly oriented $\Omega=0$.

At a volume fraction $\phi=0.4$, the equilibrium triblock system does not exhibit hexagonal order for any temperature $T$ and there is no global alignment as shown in Figure \ref{fig:align}. This is contrasted with the active system where at high temperatures and sufficiently large propelling speeds (e.g., $U=6$ or $U=8$) a significant degree of orientational order develops as signified by the large values of $\Omega$ in Figure \ref{fig:align}. It is particularly revealing to look at the data for $U=8$, as at these activities a large degree of hexagonal order is observed across the different temperatures ($>70\%$), but the local alignment of the propelling axes only occurs for large values of $T$. Strikingly, the orientational order tends to develop mostly in the outer shells of the crystal, where large ordered domains having an average orientation pointing toward the center of the hexagonal crystal are clearly visible (See Figure \ref{fig:align_snapshot}b), while its core remains disordered. An equally important observation is that no net orientational order is observed for triblock colloids within a kagome crystals. This observation appears to be true for both the active and equilibrium triblock system, and indicates that there is no mechanism at play that drives orientational alignment of colloids within the kagome lattice and that colloids are incorporated into the structure at random. 
 
These results related to the orientational alignment provide insight into the mechanisms behind the active destabilization of the kagome lattice in favor of the hexagonal lattice. At low temperatures, unbalanced shear forces can easily develop within the orientationally disordered kagome lattice for large values of $U$. When the active forces  dominate over the adhesive forces between the particles the kagome lattice is destabilized in favor of the hexagonal crystal. This is consistent with the observation that when a hexagonal crystal develops from a kagome lattice or coexists with it, no significant degree of orientational order is observed (See Figure \ref{fig:align_snapshot}a). However, as one further increases $U$ or $T$, the disordered hexagonal structure is itself unstable against the action of the propelling forces. The only mechanism that allows for the formation of a stable structure under these conditions is to generate configurations with a net radial pressure compressing the crystal from the boundary. As illustrated in Figure \ref{fig:align_snapshot}b this requires highly aligned hexagonal domains with their net orientation directed inward toward the center of the crystal.

Lastly, we should note that in this high temperature and highly active regime, orientational alignment does not develop from within the core of an already assembled crystal structure, but rather develops early on in the assembly process as the crystal nucleates from the fluid. An initially disordered small crystallite templates its own growth fueled by the addition of colloids or short chains from the bulk. Given the active nature of the colloids, free-swimming colloids joining the crystallite will be more likely to have their propelling axis directed inward and toward the center of the cluster they join. As the crystallite grows it preferentially forms layers of colloids oriented predominately in the same direction. Since the sum of all active forces acting on these crystals is typically not equal to zero, the developing crystals are typically quite active in their own right, and they tend to rotate and move across the system with a speed that is proportional to the imbalance of active forces and torques within the crystal. It is the merging of these highly oriented crystalline domains that gives rise to the stable hexagonal crystals that result in the final configurations (See Figure \ref{fig:align_snapshot}b).

\section{Conclusion}
In this paper, we have shown how even a small amount of activity can have a profound impact on the self-assembly pathway of triblock Janus colloids intended to spontaneously organize into a macroscopic kagome lattice. We find significant yields of kagome assembly can be obtained under system conditions where no assembly is possible in the corresponding equilibrium system. Activity is able to sufficiently smooth the underlying energy landscape such that the kinetically favored intermediates can quickly transform into the thermodynamically favored product. The ability of active particles to efficiently break apart branching and competing metastable structures, while generating high density fluctuations favoring the growth of isotropic structures such as the kagome and the hexagonal crystals, is the key element responsible for the enhanced self-assembly observed in this system. 
 
Although we study this problem through the lens of a single model system, we believe that the general approach and methodology discussed here can be successfully applied to other patchy colloidal systems. For instance, we have performed several simulations with colloids carrying four orthogonal attractive patches tailored to form a colloidal square lattice \cite{Zhang2004}, and find an analogous trend where by modulating the activity of the colloids it is possible to greatly improve the self-assembly of the square lattice. It is also important to emphasize that the structural changes introduce by activity are completely reversible. Once activity is turned off, the system will relax toward its stable equilibrium configuration. In systems where activity can be dynamically modulated, this suggests a number of protocols that can be implemented to maximize self-assembly. One can think of self-propulsion in these systems as a very selective filter that only allows for the stabilization of certain structures.

\section{Supporting Information}
\noindent The Supporting Information is available free of charge on the ACS Publications website at DOI: 10.1021/jacs.8b12165. 
\\
\\
\noindent Additional simulation details, and movie captions of representative state points (PDF) \\
\\
\\
\noindent Movies S1--S4 (ZIP) 

\begin{acknowledgement}
A.C. acknowledges financial supported from the National Science Foundation under Grant No. DMR-1703873. S.A.M. acknowledges financial support from the NSF Alliance for Graduate Education and the Professoriate under Award No. 1647273. We gratefully acknowledge the support of the NVIDIA Corporation for the donation of the Titan V GPU used to carry out this work. We thank Austin Dulaney and Ahmad Omar for insightful discussions and helpful comments.

The authors declare no competing financial interest
\end{acknowledgement}

\bibliography{kagome}
\end{document}